\documentclass[structabstract]{aa}
\usepackage{natbib}
\usepackage{graphicx}
\usepackage{txfonts}
\def\arcsec{\hbox{$^{\prime\prime}$}}

\def\lya{Ly$-\alpha$~}
\def\lyb{Ly$-\beta$~}

\begin{document}
   \title{The \lya profile and center-to-limb variation of the quiet Sun}
   \author{W. Curdt\inst{1}
          \and H. Tian\inst{1, 2}
          \and L. Teriaca\inst{1}
          \and U. Sch\"uhle\inst{1}
          \and P. Lemaire\inst{3}
          }
   \institute{Max--Planck--Institut f\"ur Sonnensystemforschung,
   Max-Planck-Str. 2, 37191 Katlenburg-Lindau, Germany\\
   \email{curdt@mps.mpg.de}
\and School of Earth and Space Sciences, Peking University, China
\and Institut d'Astrophysique Spatiale, Unit\'e Mixte CNRS -- Universit\'e de
Paris XI, 91405 Orsay, France
             }
   \date{Received August 25, 2008; accepted October 22, 2008}
\abstract
   {}
   {We study the emission of the hydrogen Lyman-$\alpha$ line in the quiet Sun,
   its center-to-limb variation, and its radiance distribution. We also compare
   quasi-simultaneous \lya and \lyb line profiles.}
   {We used the high spectral and spatial resolution of the SUMER spectrometer and
   completed raster scans at various locations along the disk. For the first time, we
   used a method to reduce the incoming photon flux to a 20\%-level by partly closing the
   aperture door. We also performed a quasi-simultaneous observation of both
   \lya and \lyb at the Sun center in sit-and-stare mode. We infer the flow characteristic in
   the \lya map from variations in the calibrated $\lambda$\,1206~Si\,{\sc iii} line centroids.}
   {We present the average profile of Ly$-\alpha$, its radiance distribution,
   its center-to-limb behaviour, and the signature of flows on the line profiles.
   Little center-to-limb variation and no limb brightening are observed in the
   profiles of the \lya line. In contrast to all other lines of the Lyman
   series, which have a red-horn asymmetry, \lya has a robust and - except for dark locations - dominating
   blue-horn asymmetry. There appears to be a brightness-to-asymmetry relationship.
   A similar and even clearer trend is observed in the downflow-to-asymmetry
   relationship. This important result is consistent with predictions from models
   that include flows. However, the absence of a clear center-to-limb variation in
   the profiles may be more indicative of an isotropic field than a mainly
   radial flow.}
   {It appears that the ubiquitous hydrogen behaves in a similar way to a filter
   that dampens all signatures of the line formation by processes
   in both the chromosphere and transition region.}

   \maketitle
%
%________________________________________________________________
\section{Introduction}

\lya is by far the strongest line in the VUV spectral range and must play a
dominant role in many -- if not all -- processes  in the solar atmosphere.
Irradiance at the line center excites atomic hydrogen in the cool material
of our solar system by resonant scattering, playing an important role
in planetary, cometary, and heliospheric physics.
Despite its importance, high-resolution spectroscopic observations of this line
are scarce and fundamental information is still missing or based on indirect
methods. We present unprecedented observations of the solar emission in the
\lya line of neutral hydrogen with a detailed analysis of line profiles at
various locations on the solar disk.

Three decades ago, the calibrated line profiles over the full disk
of \lya and \lyb were obtained from {\it OSO 8} data \citep{Lemaire78}, while
the {\it Skylab} ATM \citep{Reeves76} provided moderate resolution radiance maps
at different locations on the disk.
\citet{Basri78} used \lya rocket flight spectra obtained by the HRTS instrument
to investigate the network contrast and center-to-limb variation in the line profile.
One decade later, \citet{Fontenla88} reported observations of the \lya profile
for various solar disk features obtained by the {\it SMM}/UVSP instrument.
However, all of these measurements were hampered by the strong, narrow
geocoronal absorption at the line center.

With the advent of {\it SoHO}, which offers uninterrupted and undisturbed observations
of the Sun from its orbit around the Lagrangian Point L$_1$, new attempts
were made to study this line with unprecedented spectral resolution.
\citet{Lemaire98} could deduce the average line profile of the solar irradiance
by observing off disk the scattered light from the primary mirror of the SUMER
spectrometer. They found a \lya profile that had nearly no asymmetry.
\citet{Lemaire02} compared the profiles of \lya and \lyb and report calibrated
irradiances with 10\% uncertainty. A few years later, \citet{Lemaire05}
investigated the variation in \lya during solar cycle 23.
In contrast to Ly$-\alpha$, which is generally too strong to be observed on disk by SUMER, \lyb
and the higher members of the Lyman series were repeatedly observed. \citet{Warren98} completed a
comprehensive analysis of these profiles. They found that
(1) all lines from levels with main quantum number $n\ge 2$ are self-reversed
and have a shape that deviates substantially from
Gaussian with a self-reversal that decreases with increasing $n$;
(2) all lines are asymmetric, the red horn being higher than the blue horn;
(3) interestingly, the red-horn asymmetry is more pronounced for the brighter network areas.

\citet{Emerich05} established a new relationship between the line center
H\,{\sc i}~\lya irradiance and the total line irradiance derived from the scattered light
from the primary mirror of the SUMER instrument. % on {\it SoHO} during the solar cycle 23.
This work superseded earlier work of \citet{Vidal-Madjar80}, which used {\it OSO 5}
data and where the \lya line had to be corrected for geocoronal absorption.

\citet{Fontenla02} published a model to describe the \lya line formation. They
demonstrated that static models do not reproduce the observations and more reliable results can
be obtained by assuming that flows and diffusion play a major role.

\begin{figure*}
   \centering
   \sidecaption
   \includegraphics[width=11.8cm]{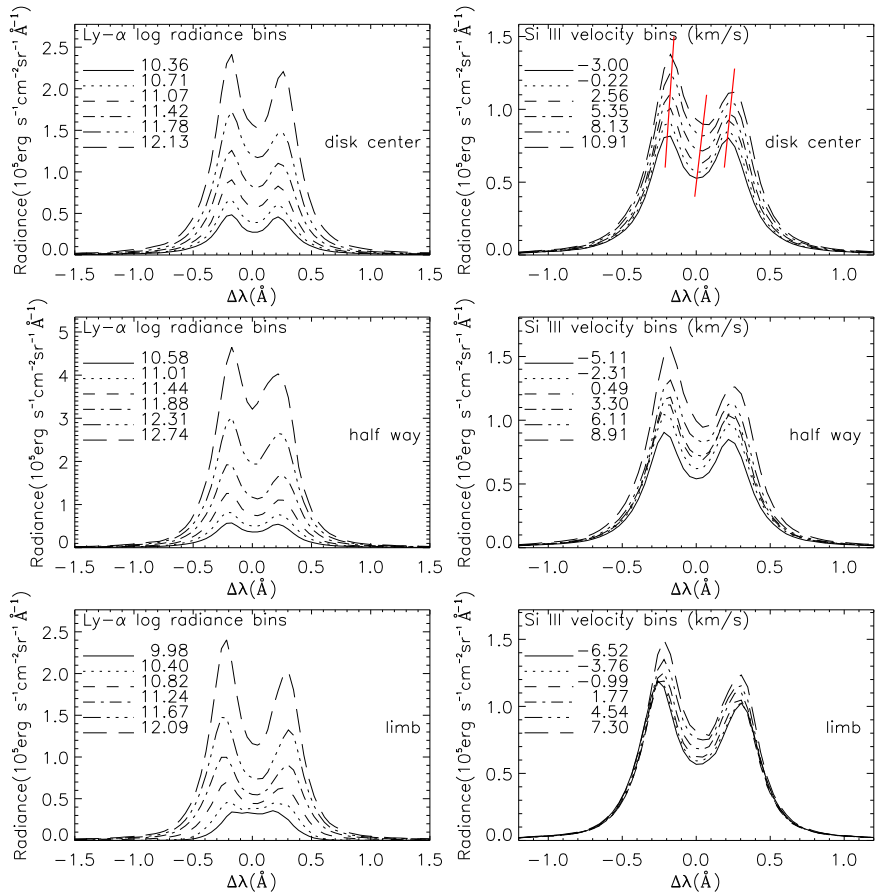}
      \caption{{\bf Left:} \lya profiles of the 26 June data set for six different radiance bins
      obtained from all data points of raster scans at disk center, at a position close to the east limb,
      and at a position half-way to the limb. In logarithmic representation,
      the bins of log radiance / s cm$^2$ sr {\AA} erg$^{-1}$  are
      equally spaced.\newline
      {\bf Right:} The Doppler flow in km/s$^{-1}$
      for each pixel was established by the deviation
      of the $\lambda$\,1206~Si\,{\sc iii} line centroid from the rest wavelength
      and six equally spaced velocity bins were defined. Negative values
      correspond to upflows, and positive values to downflows. Here, we have
      excluded data points obtained off-disk.
      The different \lya profiles for these six velocity bins are
      shown.\newline
      The blue horn is always stronger than the red horn. The asymmetry
      increases significantly with the redshift. Near disk center, all line features
      (red horn, blue horn, central depresssion) are offset towards longer wavelenths
      with increasing downflows, as indicated by the red tracing lines. Close to the limb, the velocity bins
      are set up almost symmetrically relative to rest, and the variation in the radiance
      with velocity is significantly lower than on disk.
      This effect is expected for mostly radial plasma downflows in optically thin TR emission
      such as $\lambda$\,1206~Si\,{\sc iii}.
      }
\label{bins}
\end{figure*}

%__________________________________________________________________
\section{Method}
SUMER can observe both the entire Lyman series and the wavelength range from 666~{\AA} to the Lyman limit,
which is dominated by the hydrogen recombination continuum. Due to its
enormous brightness, the \lya line, however, inundates the SUMER detectors
during observations. The high photon rate produces a gain depression in the
microchannel plates (MCP) such that single pulses cannot be detected.
To overcome this shortcoming, 1:10 attenuating grids above 50 pixels on both sides of
the detectors were introduced into the optical design of the instrument \citep{Wilhelm95}.
It became clear, however, that the attenuation also exerts a modulation on
the line profile, which makes it difficult to interpret these data, and
the attenuators have therefore not been used here.

Several attempts were made to observe this line on the bare
part of the detector, and to extrapolate the gain-depression correction
to the high photon input rate attained during those exposures. However,
this introduced a high uncertainty in the signal determination,
in particular for the brighter pixel locations \citep{Teriaca5a,Teriaca5b,Teriaca06}.

For the first time, a new method and non-routine operation were applied, which
employed the aperture door to reduce the incoming photon flux to a moderate level.
The partial closing of the door -- accomplished during a real-time commanding session prior to
the observation and re-opening thereafter -- reduced the
input photon rate by a factor of $\approx5$.
As a prologue for these observations, full-detector images in the Lyman continuum around
{880~\AA} were obtained with open and partially-closed door.
At this wavelength setting, the illumination of the detector is rather
uniform, and no saturation effects had to be considered. In this way,
accurate values for the photon flux reduction could be established.
To date, four \lya observations with partially-closed door have been performed.

\section{Observations}

On June 24 and 25, 2008, we obtained 120\arcsec~$\times$~120{\arcsec} rasters at the disk center, at
a location close to the east limb, and at a position half-way to the limb in the equatorial plane.
The limb scan extended to a position 10{\arcsec} above the limb.
Two spectral windows were transmitted, 100~pixels (px) around Ly$-\alpha$ recorded on the bare
photocathode of the detector, and 50~px around $\lambda$\,1206~Si\,{\sc iii} recorded
on the KBr part of the photocathode, respectively. All observations were completed with
the 0.3\arcsec~$\times$~120{\arcsec} slit. With an exposure time of 15~s, both lines
were observed with sufficient counts for a good line profile analysis.

A similar observation along the central meridian and including the South Pole
was performed on June 26, 2008. On July 2, 2008, a sit-and-stare sequence was
completed at disk center with a second wavelength setting for \lyb. We monitored a projected area of 100~px around
\lyb on the KBr portion of the MCP and 50~px around
$\lambda$\,1032~O\,{\sc vi} on the bare MCP. Each time, the back-and-forth movement of the
wavelength mechanism between subsequent exposures lasted for $\approx10$~seconds and
led to a cadence of 50.5~s (15\,s~+~10.25\,s~+~15\,s~+~10.25\,s).
The eastern hemisphere of the solar disk was void of active regions on all four days.

All four data sets were processed with standard procedures of the
SUMER-soft library. To complete the flatfield correction, we used a dedicated
flatfield exposure, which was acquired close in time on June 28, 2008.
The flatfield-correction matrix was obtained by applying a 16~px $\times$ 16~px
median filter to the flatfield exposure.

\section{The \lya profile and radiance distribution}

We sorted the pixels of all observed locations on the disk by the total line radiance and
defined six equally spaced radiance bins. The profiles for these bins are
displayed in the left portion of Fig.\ref{bins},
complementing the work of \citet{Warren98}. All profiles are self-reversed, show a blue-horn
asymmetry and have a similar shape, independently of brightness and limb-distance.
There is, however, a clear
brightness-to-asymmetry relationship. Brighter features show a more asymmetric profile.

We now produce a Doppler map using the simultaneously recorded $\lambda$\,1206~Si\,{\sc iii} line.
We assume that the Si\,{\sc iii} emission observed in off-disk positions was free
from systematic flows, which allowed us -- after a correction of 1.8 km/s for the solar
rotation -- to establish a rest wavelength for Si\,{\sc iii}. The rest wavelength may still have
been affected by spicular activity above the limb. It would have been
impossible, however, to achieve such an accurate absolute wavelength
calibration from the self-reversed \lya line. In a similar way to the radiance case, we
now define equally spaced velocity bins reaching from -3.0~km/s to 10.9~km/s
for the scan at disk center. As displayed in the right section of
Fig.\ref{bins}, this result is consistent with the well-known net redshift
of TR emission \citep[e.g.,][]{Brekke97,Teriaca99}. The closer the approach to the limb,
the more uniform the line strengths are when sorted into velocity bins; here, the velocities
of the bins are almost symmetric relative to zero, an effect expected for radial downflows.
It is obvious that the profiles, which are almost symmetric for locations with
little or no downflow, but become increasingly asymmetric with increasing downflow.
This result confirms and extends the observations of \citet{Fontenla88}.
It is also interesting to see that all line features (red horn, blue horn, and
central depression) are offset towards longer wavelenths with increasing downflow
velocities. This is a clear indication that, at locations with downflows, the absorption in the red
horn is increased, and that flows play a major role. These findings are consistent
with the downflow model of \citet{Fontenla02}.

\begin{figure}
   \centering
   \includegraphics[width=8cm]{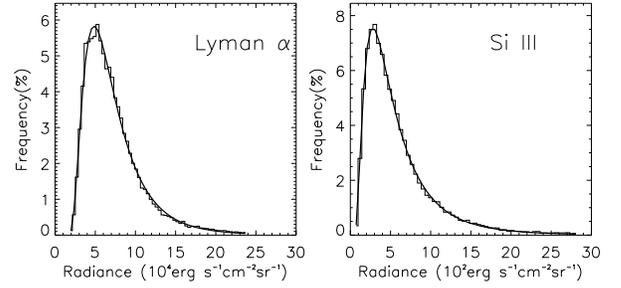}
      \caption{Radiance histogram of the raster scans in \lya emission (left) and
      $\lambda$\,1206~Si\,{\sc iii} emission (right) compared
      with a single log-normal distribution function.}
\label{histo}
\end{figure}

Motivated by the similarity between the results to date, we accumulated the
pixels of all disk center and half-way scans to derive a more robust statistic. The
distributions of the full line radiances, $x$, of both \lya and $\lambda$\,1206~Si\,{\sc iii}
are shown in Fig.~\ref{histo}. They almost perfectly follow a log-normal distribution\\\\
f($x$)~=~$\frac{a_0}{x - a_3} \cdot \exp\,(\frac{-(\ln\,(x-a_3)-a_1)^2} {2
a_2^2})$\hspace{8mm} with coefficients\\\\
$(a_0, a_1, a_2, a_3)_{Ly-\alpha}$ = (24.79, 1.623, 0.591, 1.284) \newline
and~~$(a_0, a_1, a_2, a_3)_{Si III}$ = (24.54, 1.457, 0.746, 0.338).

\vspace{2mm}\citet{Fontenla88} obtained a similar but noisier result from UVSP data. Our
result complements the work of \citet{Pauluhn00}, which was focussed on optically thin
TR lines.

%\pagebreak

\section{The \lya center-to-limb variation}

\begin{figure}
   \centering
   \includegraphics[width=8cm]{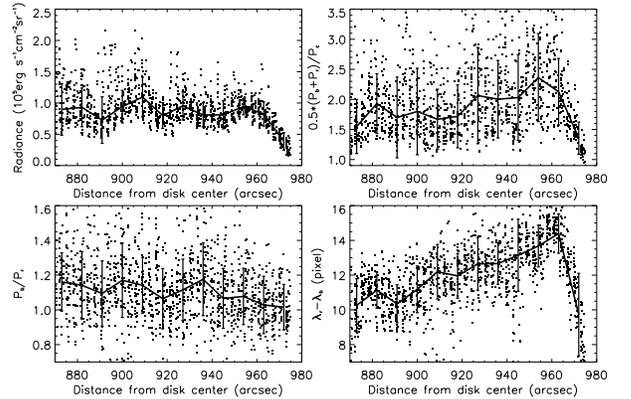}
      \caption{The center-to-limb variation in the 26 June near-limb data set.
      From the polar scan, we show the line radiance (upper left), the
      construct $0.5\cdot (P_b + P_r)/P_c$ (upper right), the ratio $P_b / P_r$ (lower left),
      and the horn separation $\lambda_r - \lambda_b$ (lower right).
      The solar radius to the photospheric limb was 954{\arcsec} as seen from {\it SoHO}.}
\label{CLV}
\end{figure}

For all observed locations, we determined descriptive parameters for the profile,
namely the peak radiances of both horns, the minimum radiance of the central depression
$P_b$, $P_r$ and $P_c$, and the corresponding
wavelengths $\lambda_b$, $\lambda_r$, and $\lambda_c$. Polynomial fits were
applied to achieve values with sub-pixel resolution.
The scatter plots for these parameters versus limb distance (not shown here)
are rather flat, since no parameter shows a significant trend. However, in the near-limb
raster of the data set on 26 June, which includes the southern polar coronal hole,
a clear trend is seen in the horn separation $\lambda_r - \lambda_b$ (lower right
in Fig.\ref{CLV}) and to a lesser extent in the construct $0.5 \cdot (P_b + P_r)/P_c$
(upper right). The trend in the ratio $P_b / P_r$ (lower left) is less significant
still and the full line radiance (upper left)
shows no trend. More observations of this type are needed to substantiate this
result.

Our radiance data also allows us to test the relation of the central \lya spectral
irradiance and the total \lya line irradiance published by \citet{Emerich05}.
The scatter plots (not shown here) suggest an almost linear relationship with
significant center-to-limb variation. We found ratios of line center radiance
to full line radiance of 9.35~$\pm$~0.24 at disk center, 9.25~$\pm$~0.10 half-way
to the limb, and 5.91~$\pm$~0.18 at the limb. This
suggests a deeper self-reversal close to the limb.

\begin{figure*}
 %  \centering
   \sidecaption
   \includegraphics[width=11.8cm]{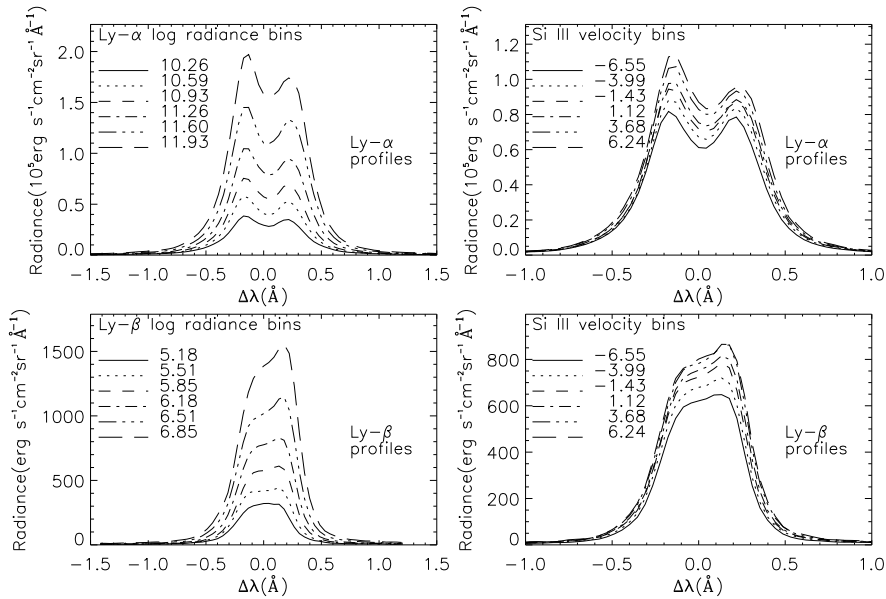}
      \caption{Comparison of quasi-simultaneous line profiles of \lya and \lyb
      obtained in sit-and-stare mode at disk center. The elapsed time between
      the exposure start of consecutive frames was only 25.3~s, and we assume only
      minor temporal variations of the emitting plasma.
      We show the profiles for different radiance bins (left) and for different
      velocity bins (right). In logarithmic representation, radiance bins are equally
      spaced. Velocity bins are defined in the same way as in  Fig.\ref{bins},
      i.e. by a pixel-by-pixel comparison of the Si\,{\sc iii} line centroids
      with the rest wavelength. Negative values correspond to upflows,
      positive values to downflows.\newline
      It is most obvious that the asymmetries in the \lya and \lyb lines are
      reversed, and there is a clear correspondence between asymmetry and
      downflows for both lines.
      }
\label{ab_bins}
\end{figure*}

Our result is in agreement with \citet{Emerich05} for a linear fit in their Fig.~6.
%Of course, spectra from the disk have a much higher dynamic range than disk-average spectra.

\section{\lya and \lyb comparison}

To complete a pixel-by-pixel comparison of both lines we had to compensate for the
small offset along the slit introduced by the movement of the wavelength
mechanism. This was achieved by the {\it delta\_pixel} routine of the SUMER-soft library. The data set
was analyzed in similar steps as described in Sect.~4.
The results of our analysis of both the \lya and \lyb line profiles are presented in
Fig.\ref{ab_bins}. In a similar way to Fig.\ref{bins}, we show the profiles for various radiance bins and
Dopplershift bins. Although there is an obvious correspondence between the asymmetry and downflows of both lines,
it is also clear that the asymmetries in the \lya and \lyb lines are reversed.
This last, unexpected result is reproduced poorly by the model with
the stronger downflow in \citet{Fontenla02}. However, the model
(as discussed also by its authors) does not reproduce well the observed
profiles (particularly that of \lyb), indicating that significant improvement
is still required in modeling these lines and the solar structures in which they originate.

The relationship between asymmetry and downflows and the comparison with models
would, thus, indicate the presence of a persistent downflow such as that observed
in the optically thin TR lines. The apparent lack of a center-to-limb variation
in the line asymmetry is consistent with a quite isotropic system of downflows.
In the case of optically thin lines, the opposite flows cancel out at the limb
as indicated by the disappearance of the average downflow of TR lines at the limb
\citep{Peter99}, while this is not necessarily true for the thick H\,{\sc i} Lyman lines.

To characterise the line ratio, we prepared a scatter plot of the \lya line radiances versus \lyb line
radiances (not shown here). We found an almost linear relationship with a
photon per line ratio of 188 $\pm$ 1. Our value is considerably higher than
those measured to be 121 and 85 by \citet{Vernazza78} and \citet{Lemaire05}, respectively.
The numerical uncertainty in our result is negligible if compared with other error sources.
For the uncertainty introduced by the relative radiometry, we estimate a conservative
value of 10\% (the sensitivity curve of SUMER is rather flat in the range from
1020~{\AA} to 1220~{\AA}). The discrepancy between this new result and previous
observations of ATM and OSO-8 may be due, at least in part, to the poorer quality of the photometry
used in these earlier reports. It may also be because the selection of only
120 pixels provided a sample that was unrepresentative of the mean behavior of the Sun,
and may therefore be resolved by completing similar observations but with a
larger raster in both lines. We have initiated  a dedicated project to investigate this issue further.

\section{Summary and conclusion}
We have presented the first on-disk SUMER observations of \lya with reduced
incoming photon flux and some fundamental results derived thereof.
Such observations are important in many respects as already
demonstrated here. It is, however clear that this work can only
be the starting point of greater efforts to understand the excitation and absorption
processes responsible for the line formation, and eventually to
constrain atmospheric models. From our work, we conclude that downflows
must play a fundamental role in line profile formation. We plan further
observations during the rise of solar cycle 24. These will reveal more details
and indicate, whether there is an imprint of solar activity on the profile
of this important line.

\begin{acknowledgements}
The SUMER project is financially supported by DLR, CNES, NASA, and the ESA PRODEX
Programme (Swiss contribution). SUMER is part of {\it SOHO} of ESA and NASA.
HT is supported by China Scholarship Council for his stay at MPS. This
non-routine observation was performed with the help of D. Germerott. We
thank V. Andretta and an anonymous referee for valuable comments and suggestions.

\end{acknowledgements}

\end{document}